\documentclass[useAMS,usenatbib,referee]{mn2e}
\usepackage{graphicx, amssymb}

\begin{document}

%
%

\title[Analysing lightcurves of eclipsing binaries in LMC with EBAS]
{Automated analysis of eclipsing binary lightcurves with EBAS. II.
Statistical analysis of OGLE LMC eclipsing binaries}

\author[T. Mazeh, O. Tamuz \& P. North]{ T.~Mazeh,$^1$\thanks{E-mail:
mazeh@wise.tau.ac.il} O.~Tamuz$^1$ and P.~North$^2$\\ $^1$School of
Physics and Astronomy, Raymond and Beverly Sackler Faculty of Exact
Sciences, Tel Aviv University, Tel Aviv, Israel\\ $^2$Ecole
Polytechnique F\'ed\'erale de Lausanne (EPFL), Laboratoire
d'Astrophysique,\\
Observatoire, CH-1290 Sauverny,
Switzerland}

\maketitle


\begin{abstract}

In the first paper of this series we presented EBAS, a new fully
automated algorithm to analyse the lightcurves of eclipsing binaries,
based on the EBOP code. Here we apply the new algorithm to the whole
sample of 2580 binaries found in the OGLE LMC photometric survey and
derive the orbital elements for 1931 systems. To obtain the
statistical properties of the short-period binaries of the LMC we
construct a well defined subsample of 938 eclipsing binaries with
main-sequence B-type primaries. Correcting for observational selection
effects, we derive the distributions of the fractional radii of the
two components and their sum, the brightness ratios and the periods of
the short-period binaries. Somewhat surprisingly, the results are
consistent with a flat distribution in log P between 2 and 10 days. We
also estimate the total number of binaries in the LMC with the same
characteristics, and not only the eclipsing binaries, to be about
5000. This figure leads us to suggest that $(0.7\pm 0.4)$\% of the
main-sequence B-type stars in the LMC are found in binaries with
periods shorter than 10 days. This frequency is substantially smaller
than the fraction of binaries found by small Galactic radial-velocity
surveys of B stars. On the other hand, the binary frequency found by
HST photometric searches within the late main-sequence stars of 47 Tuc
is only slightly higher and still consistent with the frequency we
deduced for the B stars in the LMC.

\end{abstract}


\section{Introduction}

A number of large photometric surveys have recently produced
unprecedentedly large sets of high S/N stellar lightcurves
\citep[e.g.,][]{alcocketal97}. One of these projects is the OGLE study
of the SMC \citep{udalski1998} and the LMC \citep{udalski2000}, which
has already yielded \citep{lukas2003} a few thousand lightcurves of
eclipsing binaries. These two sets of lightcurves enable a statistical
analysis of a large sample of short-period extragalactic binaries, for
which the absolute magnitudes of all systems are known to a few
percent. No such sample is known in our Galaxy.

One such statistical study was the analysis of \citet[][]{NZ03}, who
derived the orbital elements and stellar parameters of 153 eclipsing
binaries discovered by OGLE in the SMC.  North \& Zahn examined the
eccentricities of the eclipsing binaries as a function of the ratio
between the stellar radii and the binary separation, and compared
their result with a similar dependence derived from the elements of
the binaries discovered by \citet{alcocketal97} in the LMC. They then
considered the implication of their findings for the theory of tidal
circularization of early-type primaries in close binaries
\citep[e.g.,][]{zahn75, zahn77, ZB89}.

In a follow-up paper \citet{NZ04} analyzed another
set of 510 lightcurves selected from the 2580 eclipsing binaries
discovered in the LMC by the OGLE team \citep[][]{lukas2003}.  Again,
they derived the ratio between the stellar radii and the binary
separation for these systems, and examined the dependence of the
binary eccentricity on this ratio.

The previous analyses used only a small fraction of the sample of
eclipsing binaries found in the OGLE data. The goal of the present
study is to analyze the whole sample of 2580 lightcurves discovered by
OGLE in the LMC, and to derive some statistical properties of the
short-period binaries, after correcting for observational selection
effects. Such a correction is essential for the derivation of the
period distribution, for example, because the probability of detecting
an eclipsing binary is a strong function of the orbital period. In
order to apply an appropriate correction one needs a complete
homogeneous data set of lightcurves, all discovered by the same
photometric survey of a well defined sample of stars. These
requirements were exactly met by the OGLE data set for the LMC,
enabling such analysis for a large sample of eclipsing binaries for
the first time.

A completely automated algorithm is needed to analyse the large set of
eclipsing binaries at hand. Two such codes were developed
recently. \citet{WW1} have constructed an automatic scheme
based on the Wilson-Divenney (=WD) code to analyze the OGLE
lightcurves detected in the SMC in order to find eclipsing binaries
suitable for distance measurements.  At the last stages of writing
this paper another study with an automated lightcurve fitter ---
DEBiL, was published \citep{devor2005}.  DEBiL was constructed to be
quick and simple, and therefore has its own lightcurve generator,
which does not account for stellar deformation and reflection
effects. This makes it especially suitable for detached binaries. We
will use in this work our EBAS, which was presented in Paper I (Tamuz,
Mazeh \& North 2005) and is based on the EBOP code.  The
complexity of EBAS is in between DEBiL and the automated WD code of
Wyithe \& Wilson.

To facilitate the search for global minima in the convolved parameter
space, EBAS performs two parameter transformations. Instead of the
radii of the two stellar components of the binary system, measured in
terms of the binary separation, EBAS uses the
total radius, which is the sum of the two relative radii, and their
ratio. Instead of the inclination we use the impact parameter --- the
projected distance between the centres of the two stars in the middle
of the primary eclipse, measured in terms of the total radius.

The set of parameters of the EBAS version used here includes the
bolometric reflection of the two stars, $A_p$ and $A_s$. When $A_p=1$
the primary star reflects all the light cast on it by the
secondary. Together with the tidal distortion of the two components,
which is mainly determined by the mass ratio of the two stars, the
reflection coefficients $A_p$ and $A_s$ determine the light
variability of the system outside the eclipses. Paper I discussed the
reliability of the values of these two parameters as found by EBAS.

To simplify the analysis, the present version of EBAS assumes there is
no contribution of light from a third star and that the mass ratio is
unity.  Paper I discusses the implication of these choices
on the parameters' values, showing that the values of the total radius
and the surface brightness ratio are only slightly modified by these
two assumptions. The only parameter which is systematically modified
by the third-light assumption is the impact parameter, which is
directly associated with the orbital inclination of the binary.  Note,
however, that we are not interested in one specific system but aim,
instead, at deriving the gross characteristics of the short-period
binaries.  As the analysis of this paper does not use the inclinations
of the eclipsing binaries for the derivation of the statistical
features of the short-period binaries, we regard the resulting
distributions as probably correct.

Paper I introduces a new 'alarm' statistic, $\mathcal{A}$, to replace
human inspection of the residuals.  EBAS uses the new statistic, which
is sensitive to the correlation between neighbouring residuals to
decide automatically whether a solution is satisfactory. In this work
we consider only the 1931 systems that yielded solutions with low enough
alarm value.

To check the reliability of our results we compare our geometrical
elements with those derived by \citet[][hereafter MiP05]{michalska05}
with the WD code for 85 binaries, based on EROS, MACHO and OGLE
data. The comparison is reassuring, as it shows that the geometrical
parameters of EBAS are close to the ones derived by the more
sophisticated WD approach.

To derive the orbital distributions of the short-period binaries we
had to trim the sample, to get a homogeneous sample which we
could correct for selection effects. We were left with 938
main-sequence binaries with periods shorter than 10 days and system
magnitudes between 17 and 19 in the $I$ band, most of which are
binaries with B-type primaries. This makes the range of the derived
period distribution quite narrow.  Nevertheless, the data yielded
somewhat surprising distributions, which might have implications on
binary population studies.

Section~\ref{elements} presents the resulting orbital elements of the
OGLE LMC eclipsing binaries and compares the derived elements with
those of MiP05. Section~\ref{trimming} details the procedure to focus
on a well defined homogeneous subsample and Section~\ref{statistics}
derives the statistical features of the short-period binaries, after
correcting for the observational selection
effects. Section~\ref{discussion} discusses the new findings, and
Section~\ref{summary} summarizes the paper.

\section{Analysis of the OGLE LMC Eclipsing Binaries}   
\label{elements}

The LMC OGLE-II photometric campaign \citep{udalski2000} was carried
out from 1997 to 2000, during which between 260 and 512 measurements
in the $I$ band were taken for 21 fields \citep{zebrun2001}.
\citet{lukas2003} searched the photometric data base and identified
2580 binaries. We analysed those systems with EBAS and found 1931
acceptable solutions. Three types of binaries were excluded:

\begin{itemize}

\item
Four binaries were found to appear twice in the list of binaries, with
the same period (except for a factor of 2) and with very close
positions. Apparently, they originated from the overlap between
the different fields of OGLE, and evaded the scrutiny of
\citet{lukas2003}.

\item
EBAS found 376 solutions with alarm too high --- $\mathcal A >
0.5$. Visual inspection showed that most of these systems are contact
binaries that EBOP can not model properly. Some might be ellipsoidal
variables \citep[see][]{lukas2003} or other types of periodic
variables, other than eclipsing binaries.

\item
EBAS found 269 solutions that yielded sum of radii
$r_t=r_p+r_s$ too large.
Following the Roche lobe radius calculation in \citet{Egg1983}:
\begin{equation}
R_{RL}/a=\frac{0.49\,q^{2/3}}{0.6\, q^{2/3}+\ln(1+q^{1/3})}\; \; ,
\end{equation}
which reduces for $q=1$ to $R_{RL}/a=0.379$, we did not
accept solutions with

\begin{equation}
r_t > 0.65 \, (1-e\cos \omega) \ ,
\end{equation}
assuming the EBOP model can not properly account for the deformation
of the two stars if one or two stars are too close to their Roche-lobe
limit, at least at the periastron passage.

\end{itemize}

We were thus left with 1931 acceptable solutions.

\subsection{Solution examples}

\begin{figure*}
 \includegraphics{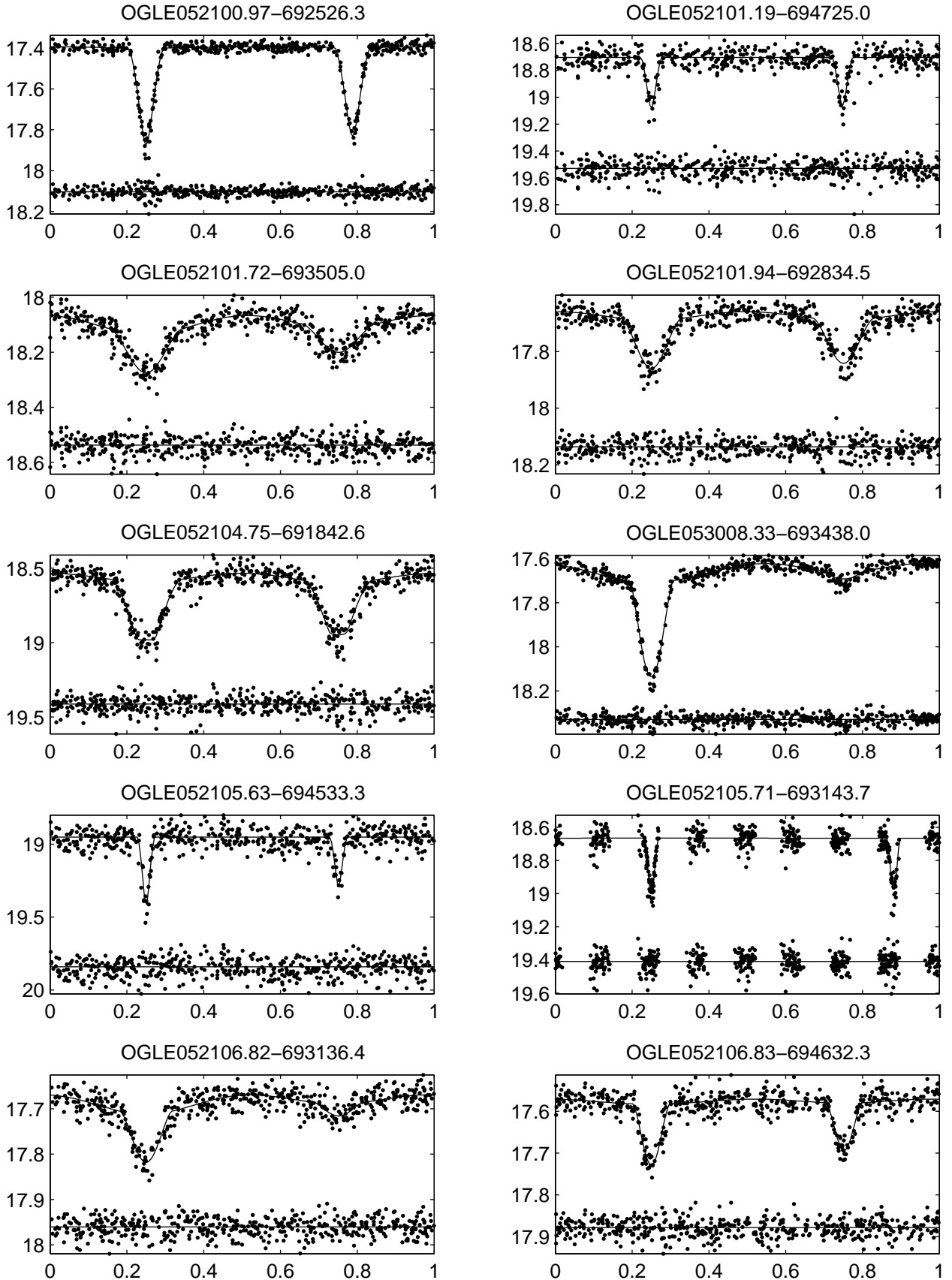}
 \caption{Ten derived lightcurves}
 \label{fig:ten_lightcurves}
\end{figure*}

In Fig~\ref{fig:ten_lightcurves} we plot 10 representative lightcurves
derived by the EBAS. Their elements are given in
Table~\ref{table:ten_binaries} and
Table~\ref{table:ten_binaries_elements}. Table~\ref{table:ten_binaries}
gives some global information on each lightcurve and the
goodness-of-fit of its model. It lists the number of measurements, the
averaged observed $I$-magnitude and the rms of the scatter of the OGLE
lightcurve. It also lists the $\chi^2$, the alarm $\mathcal A$ of the
fit and the detectability measure $\mathcal D$ (see below), and
whether this binary was included in the trimmed sample (+), as
detailed in the next section. Table~\ref{table:ten_binaries_elements}
lists the derived elements of EBAS and their estimated
uncertainties. This includes the $I$-magnitude of the system at quadrature
(the $SFACT$ parameter of EBOP, see Table~2 of Paper~I), the period, $P$,
in days, the sum of radii, $r_t$, the ratio of
radii, $k$, the surface brightness ratio, $J_s$, the impact parameter,
$x$, and the eccentricity $e$, multiplied by $\cos \omega$ and
$\sin\omega$, where $\omega$ is the longitude of periastron. Note that
the mag parameter in Table~\ref{table:ten_binaries_elements} is the
brightness of the system out of eclipse, and therefore is different
from the observed averaged mag given in
Table~\ref{table:ten_binaries}. Since its formal error is very small, we
give it to 4 or 5 decimal figures because this might be useful for comparison
purposes, even though we are well aware that systematic errors --- whether
in the original photometric data or in their fit --- are far larger than that
level of accuracy. Let us recall that some parameters were held fixed in
the fit, as explained in Paper~I: the linear limb-darkening coefficients,
with a value typical of main-sequence B stars ($u_p=u_s=0.18$), the gravity
darkening coefficients ($y_p=y_s=0.36$), the mass ratio ($q=1$), the tidal
lead/lag angle ($t=0$) and the third light ($L_3=0$). $A_p$ and $A_s$ are the
bolometric reflection coefficients, the value of which is between $0$ and $1$;
they determine (together with the tidal distorsion of the components) the
out-of-eclipse variability. They are adjusted in order to fit the variation
outside eclipses, but since reflection effects are only crudely modeled by
EBOP, one has to keep in mind that their value may have very limited physical
significance.

\begin{table*}
\centering
\begin{minipage}{140mm}
\caption{The ten binaries: observations and goodness-of-fit}
\label{table:ten_binaries}
\begin{tabular}{@{}lrrrrrrc@{}}
OGLE Name & $N$ \ & $mag$ & $rms$ & $\chi^2$ & $\mathcal A$ \ & $\log \mathcal
D$& Sample\\
\hline
052100.97-692526.3&482& 17.44&  0.12&   397&-0.3&15015&+\\
052101.19-694725.0&478& 18.73&  0.09&   453&-0.2&  688&+\\
052101.72-693505.0&479& 18.12&  0.07&   381&0.1& 1606\\
052101.94-692834.5&472& 17.70&  0.06&   335&-0.1& 1969&+\\
052104.75-691842.6&479& 18.64&  0.16&   425&-0.0& 3658&+\\
053008.33-693438.0&507& 17.69&  0.11&   410&0.4&12503&+\\
052105.63-694533.3&477& 18.98&  0.10&   376&0.1&  660&+\\
052105.71-693143.7&482& 18.71&  0.10&   401&-0.2& 1540&+\\
052106.82-693136.4&478& 17.69&  0.04&   299&0.0& 1353\\
052106.83-694632.3&481& 17.59&  0.04&   353&0.2& 1704&+\\
\hline
\end{tabular}
\end{minipage}
\end{table*}

\begin{table*}
\centering
\begin{minipage}{200mm}
\caption{The ten binaries: orbital elements}
\label{table:ten_binaries_elements}
\begin{tabular}{@{}lrrrrrrrrrrr@{}}
OGLE Name & $mag$ \ & $P$ \ \,\,\, \ \ & $r_t$ \ \,\ & $k$\ \,\ \ &
$J_s$ \ & $x$ \ \ & $e\cos\omega$ & $e\sin\omega$ & $A_p$ \ \, & $A_s$\ \,\\
\hline
052100.97&17.39743&3.1227063&0.2968&0.80&0.934&0.233&0.05902&0.0041&0.73&1.00000\\
\hspace{3mm}-692526.3&$\pm$0.00095&$\pm$0.0000061&$\pm$0.0038&$\pm$0.26&$\pm$0.015&$\pm$0.018&$\pm$0.00066&$\pm$0.0093&$\pm$0.18&$\pm$0.00026\\
052101.19&18.7058&9.156860&0.1788&1.16&0.97&0.339&-0.0038&0.005&0.9999&0.99993\\
\hspace{3mm}-694725.0&$\pm$0.0025&$\pm$0.000051&$\pm$0.0086&$\pm$0.37&$\pm$0.13&$\pm$0.062&$\pm$0.0014&$\pm$0.057&$\pm$0.0013&$\pm$0.00090\\
052101.72&18.0867&1.4570012&0.674&1.31&0.65&0.696&0.0016&0.004&0.75&0.916\\
\hspace{3mm}-693505.0&$\pm$0.0015&$\pm$0.0000070&$\pm$0.024&$\pm$0.26&$\pm$0.11&$\pm$0.016&$\pm$0.0042&$\pm$0.012&$\pm$0.20&$\pm$0.099\\
052101.94&17.6753&1.3626777&0.560&1.02&0.954&0.607&0.0003&0.011&1.000&0.998\\
\hspace{3mm}-692834.5&$\pm$0.0015&$\pm$0.0000025&$\pm$0.015&$\pm$0.10&$\pm$0.069&$\pm$0.019&$\pm$0.0029&$\pm$0.017&$\pm$0.013&$\pm$0.054\\
052104.75&18.5378&1.7568736&0.534&1.65&0.895&-0.035&0.0002&-0.018&0.55&0.11\\
\hspace{3mm}-691842.6&$\pm$0.0023&$\pm$0.0000041&$\pm$0.016&$\pm$0.48&$\pm$0.038&$\pm$0.094&$\pm$0.0032&$\pm$0.028&$\pm$0.31&$\pm$0.28\\
053008.33&17.6360&4.136428&0.4594&4.86&0.065&0.736&0.0006&-0.061&0.01&0.945\\
\hspace{3mm}-693438.0&$\pm$0.0010&$\pm$0.000015&$\pm$0.0092&$\pm$0.51&$\pm$0.018&$\pm$0.026&$\pm$0.0052&$\pm$0.028&$\pm$0.16&$\pm$0.099\\
052105.63&18.9528&6.7992460&0.1450&0.80&0.757&0.275&0.0018&-0.002&1.0000&0.990\\
\hspace{3mm}-694533.3&$\pm$0.0029&$\pm$0.0000035&$\pm$0.0074&$\pm$0.26&$\pm$0.076&$\pm$0.047&$\pm$0.0021&$\pm$0.066&$\pm$0.0011&$\pm$0.038\\
052105.71&18.6663&8.000289&0.1147&1.8&0.987&0.22&0.2069&-0.060&1.00&0.095\\
\hspace{3mm}-693143.7&$\pm$0.0022&$\pm$0.000029&$\pm$0.0053&$\pm$1.8&$\pm$0.089&$\pm$0.13&$\pm$0.0012&$\pm$0.042&$\pm$0.20&$\pm$0.040\\
052106.82&17.6869&1.6991553&0.591&1.20&0.362&0.780&0.0015&0.031&0.988&0.820\\
\hspace{3mm}-693136.4&$\pm$0.0011&$\pm$0.0000086&$\pm$0.017&$\pm$0.12&$\pm$0.094&$\pm$0.018&$\pm$0.0079&$\pm$0.022&$\pm$0.022&$\pm$0.087\\
052106.83&17.5775&4.796268&0.379&1.40&0.64&0.686&0.0029&-0.038&0.99&0.79\\
\hspace{3mm}-694632.3&$\pm$0.0011&$\pm$0.000024&$\pm$0.010&$\pm$0.24&$\pm$0.22&$\pm$0.035&$\pm$0.0024&$\pm$0.055&$\pm$0.15&$\pm$0.20\\
\hline
\end{tabular}
\end{minipage}
\end{table*}

Of the ten stars, the lightcurve of OGLE052101.72-693505 yielded $r_t$
too high, and the primary of OGLE052106.82-693136.4 is probably not a
main-sequence star (see below). Therefore both stars were not included
in the final statistical analysis. Because the system
053008.33-693438.0 has a very shallow secondary minimum, its ratio of
radii is poorly constrained, and it probably has undergone mass
exchange. Because of the very small surface brightness ratio, the
reflection coefficient is meaningful for the secondary but not for the
primary, which remains unaffected by the presence of its much cooler
companion. Thus the very small $A_p$ value has no real meaning.

The elements of all 1931 binaries are given in Table~3 and Table~4
with exactly the same format as in Table~\ref{table:ten_binaries} and
Table~\ref{table:ten_binaries_elements}.  Tables 3 \& 4 appear only in
electronic version of the journal in postscript format, while their
machine-readable ASCII version is available in our web
site\footnote{http://wise-obs.tau.ac.il/$\sim$omert/}.

\subsection{Comparison with the analysis of \citet{michalska05}}

In Paper I we compared the elements of four systems derived by EBAS
with those obtained by \citet{gonzalez05}. Here we wish to compare our
results with the more extensive work of MiP05 published very
recently. Using data from EROS, MACHO and OGLE, MiP05 fitted 98 LMC
binaries with the WD code.  Of those, 85 were included in the OGLE
catalogue and 81 were solved by EBAS with low $\mathcal A$.

\begin{figure*}
\includegraphics{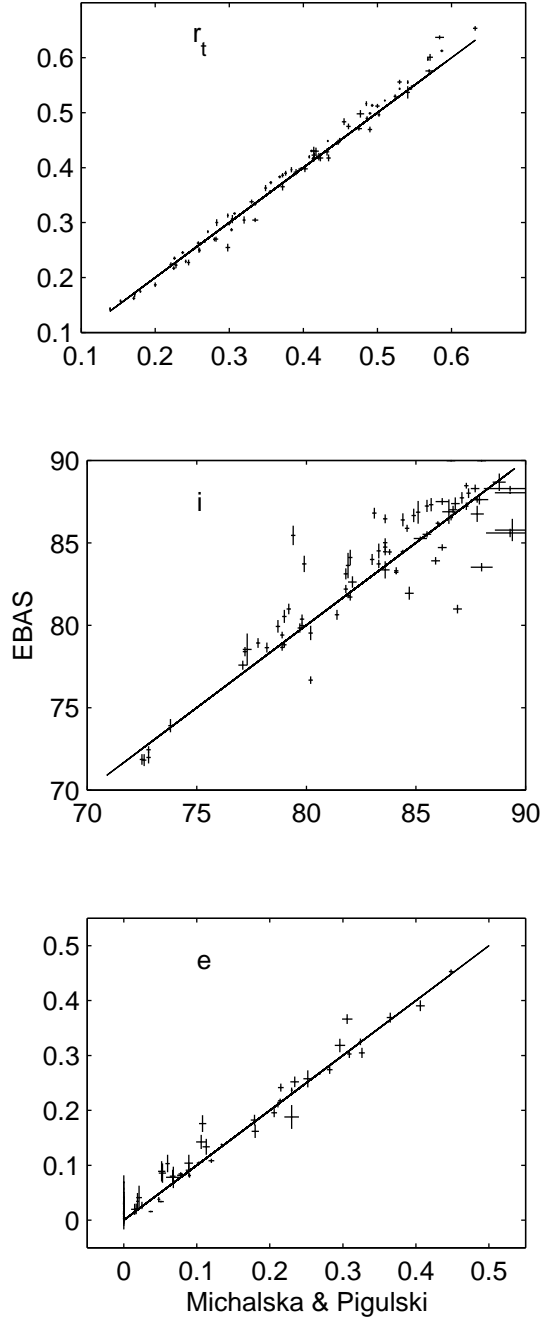}
\caption{Comparison with the work of MiP05: Elements for 81
systems. Panels are for sum of radii, inclination and eccentricity}
\label{fig:michalska}
\end{figure*}

Because of the differences between WD and EBOP, we wish to compare
only the geometric parameters of the solutions (see Paper I). Plotted
in Fig~\ref{fig:michalska} are our values for the 81 binaries versus
MiP05's, for the sum of radii, inclination and eccentricity. The total
radius and the eccentricity panels show quite small spread around the
straight lines, which represent the locus of equal values of the two
solutions. Only the inclination shows a large scatter and a slight bias.
However, as noted above, our statistical analysis does not use
the inclinations of the eclipsing binaries for the derivation of the
characteristics of the short-period binaries. Therefore, the
comparison with MiP05's solutions supports our assessment that while
individual EBAS solutions might be inferior to WD solutions, the
statistical interpretation of the entire sample remains valid.

\section{Trimming the sample}
\label{trimming}

The very large sample of 1931 short-period binaries enables us to
derive some statistical features of the population of short-period
binaries in the LMC. However, the sample suffers from serious
observational selection effects, which affected the discovery of the
eclipsing binaries. To be able to correct for the selection effects we
need a well-defined homogeneous sample. We therefore trim the sample
before deriving some parameter distributions in the
next section.

The most important observational selection effect is associated with
the weak signal of the eclipse relative to the noise of the
measurements. To apply a correction for this selection effect we need
a distinct criterion that defines the systems that could have been
detected by the OGLE LMC survey. Such a criterion is the detectability
parameter $\mathcal D$, which we define to be the number of points in
the lightcurve, $N$, times the ratio of the variance of the ideal but
variable signal, to the variance of the residuals:

\begin{equation}
\mathcal D=N\frac{var(m_i)}{var(r_i)}\ .
\end{equation}
The $m_i$'s are the values of the EBAS model at the times of
observation, and the $r_i$'s are the residuals of the measurements
relative to the model. One sees that systems with deep eclipses,
which are more easily detected, have larger $\mathcal D$, and even
more so when many measurements are concentrated within phases of
eclipses.

\begin{figure*}
\includegraphics{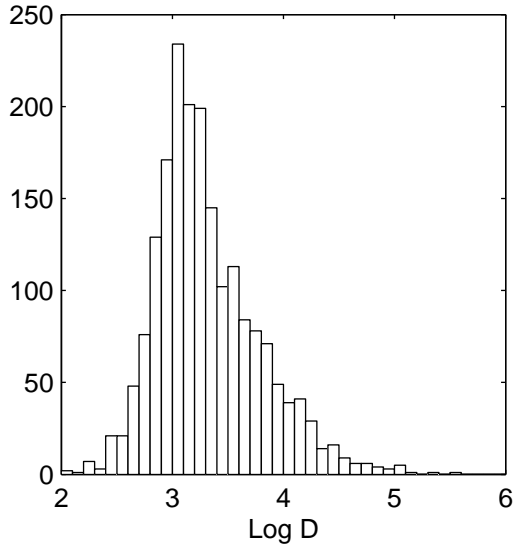}
\caption{Distribution of the detectability parameter $\log \mathcal D$
for the 1931 binaries with acceptable solutions.}
\label{fig:SN}
\end{figure*}

Fig~\ref{fig:SN} shows the distribution of $\log \mathcal D$ for all
1931 solved systems. The histogram shows that there are only very few
systems with $\log \mathcal D$ below $2.6$.  We therefore set our
detection limit, somewhat arbitrarily, at $\log \mathcal D=2.6$,
assuming that any binary below this limit could not have been
detected.  As pointed out by the referee, this limit translates into
$\mathcal D = 400$, which implies a $1\sigma$ variation in a
lightcurve having $400$ points --- a rather typical number.  We consider
the few systems below this limit as exceptions.  In order to get a
homogeneous sample we ignore the binaries below this limit, and
consider only the 1875 systems with $\log \mathcal D > 2.6$.

\begin{figure*}
\includegraphics{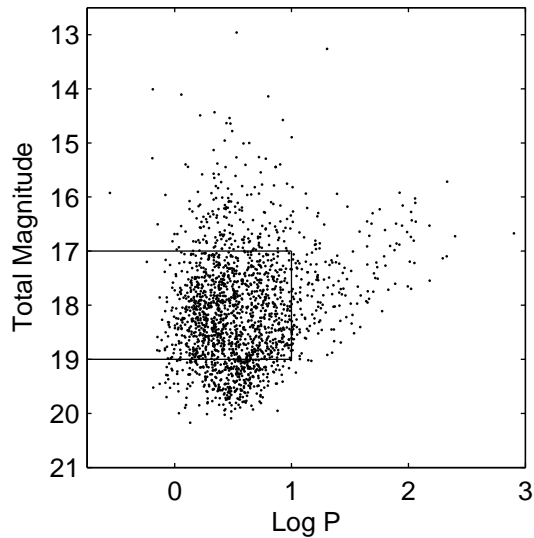}
\caption{Total $I$-magnitude vs. period for 1875 binaries with accepted
 solutions and high enough detectability. The straight lines present
 the limits of the trimmed sample. Periods are in days}
\label{fig:lum_per}
\end{figure*}

Fig~\ref{fig:lum_per} shows the total magnitude of the binaries left
in the sample as a function of their orbital periods. We are
witnessing a lack of systems in the lower right corner of the
plot. This is due to the fact that the probability of having an
eclipse is approximately equal to the sum of fractional radii,
$r_t$. The two absolute radii determine the total luminosity of
their system, while the binary separation determines the orbital
period.  Therefore, for a given period, the probability of having an
eclipse is smaller for faint systems, with smaller stellar radii, than
for brighter systems, with larger stars. In addition, the
detectability $\mathcal D$ is smaller for fainter systems which have
larger $var(r_i)$, making fainter systems less frequent in the sample.

In order to obtain a sample that does not suffer from severe
incompleteness, we chose, somewhat arbitrarily, to consider only
systems in the range of total magnitude between 17 and 19 and with
periods shorter than 10 days. The selected range, which included 1131
systems, is denoted in the figure.

For a distance modulus of $\mu_0=18.50\pm0.02$ \citep[][]{alves}, the
trimmed sample range is between absolute $I$-magnitude of $-$1.5 and
0.5, if interstellar absorption is neglected. For a metallicity
$Z=0.008$ (typical of the LMC), this range corresponds, for single
stars at ZAMS, to spectral type between B1 and B6 and mass range
between 8.7 and 3.6 $M_{\odot}$ \citep{schaerer93}. Obviously, if the
two stars have close to equal luminosities, the primary could be as
faint as 1.2 in $I$, which corresponds to a B8 spectral type.

\begin{figure*}
 \includegraphics{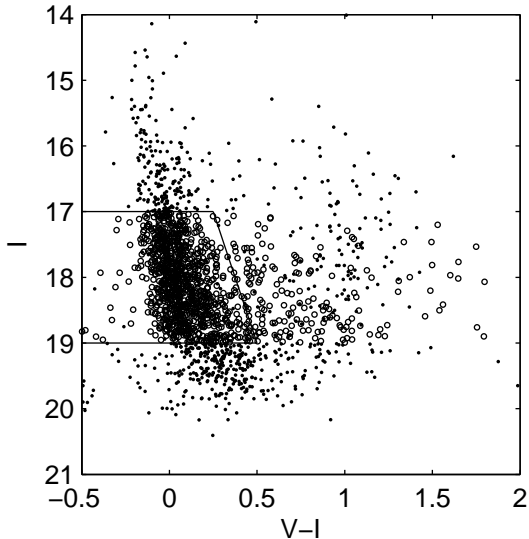}
 \caption{The $I$-magnitude as a function of the averaged $V-I$
 colour. Binaries excluded from the trimmed sample by the previous
 constraints are represented by dots. All 1131 binaries still in
 the trimmed sample are presented by small circles.}
 \label{fig:VminusI}
\end{figure*}

Finally, we calculated for all binaries the averaged $V-I$ colour
index. This was calculated by averaging all $V$ measurements that were
observed in the phase between 0.15 and 0.35 and 0.65 and 0.85, that is
out of eclipse. The $I$-magnitude of each system was taken from the
orbital solution. Fig~\ref{fig:VminusI} presents for the 1931 binaries
with reliable elements the $I$-magnitude as a function of the averaged
$V-I$ colour. Obviously, most binaries are on the main sequence, but
some binaries are not. We excluded the binaries out of the plotted
lines and were left with 938 binaries. This is the trimmed sample that
the next section uses for the statistical analysis of the short-period
binaries in the LMC.

\section{Statistical analysis of the trimmed sample}
\label{statistics}

\subsection{The radii and the surface brightness ratio distributions}

\begin{figure*}
 \includegraphics{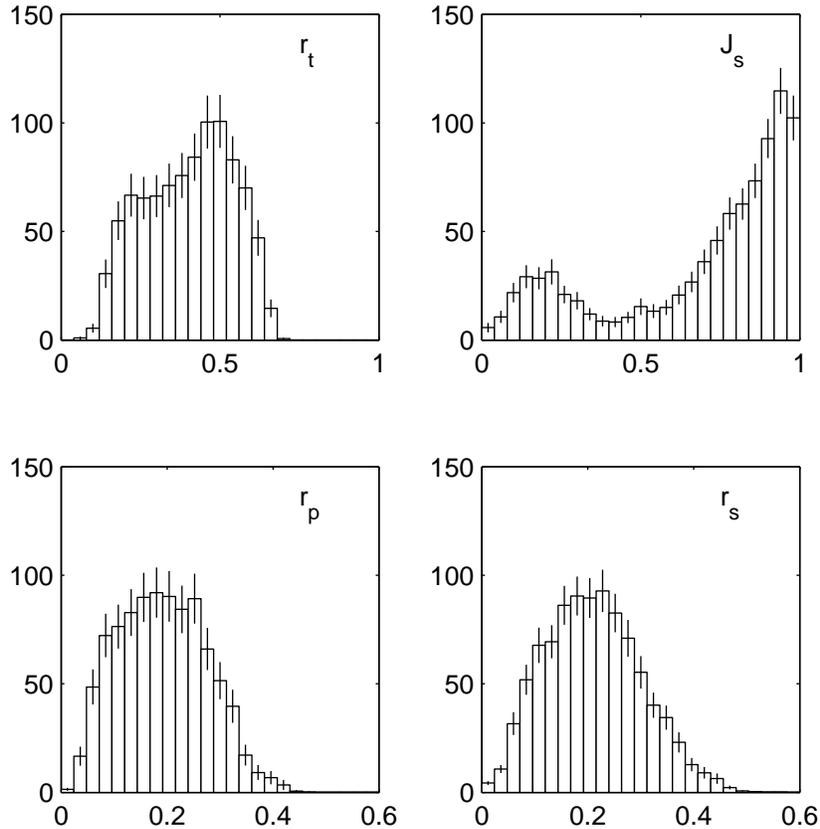}
 \caption{Histogram of the derived elements.}
 \label{fig:param_distr}
\end{figure*}

In order to study the statistical features of the short-period
binaries in the LMC with mainly B-type main-sequence primaries, we
plotted in Fig~\ref{fig:param_distr} the distribution of four orbital
elements derived for the trimmed eclipsing binary sample. This
includes the surface brightness ratio of the two stars, the
fractional primary and secondary radii, and the sum of the two radii.

The distributions are represented by histograms, which were derived by
the Gaussian kernel method \citep[e.g.,][]{silverman}. Each binary is
represented by a normal distribution centred on the derived value of
its parameter, with a width equal to the estimated uncertainty of the
parameter. Therefore, each of the histograms represents a sum of normal
distributions, differing in mean and variance. The error
bars in the histogram bins are just the square root of the value of
each bin, assuming Poisson statistics.

Obviously, the observed distributions are highly convoluted by
selection effects. For example, binary systems with longer periods
must have an inclination closer to $90^{\circ}$ to be detected as an
eclipsing binary. To correct for this effect, we calculated for each
system the minimum inclination for which the eclipse could still be
detected.  This was done by calculating for each system the minimum
inclination $i_{min}$ for which the produced eclipse would still be
above the detection threshold, given the physical parameters of the
two stellar components and the binary elements. We assumed a
lightcurve is detectable whenever $\log \mathcal D>2.6$, as above.

Assuming the systems are randomly oriented in space, the probability
of an inclination to fall between $i_{min}$ and $90^\circ$ is $p=\cos
i_{min}$.  To account for systems with inclinations outside this
range, we considered each system as representing $1/p$ binaries, and
drew the different histograms accordingly.

Actually, the probability of detection might also be a function of the
phase of the primary eclipse, especially if the data points are not
equally distributed over the binary phase and the period is close to
an integer number of days (See the discussion of \citet{pont2005} and
in particular their Figure 12). We therefore averaged each $p$ over
100 different phases of the primary eclipse, and only then assigned
each system with $1/p$ binaries. 

\begin{figure*}
\includegraphics{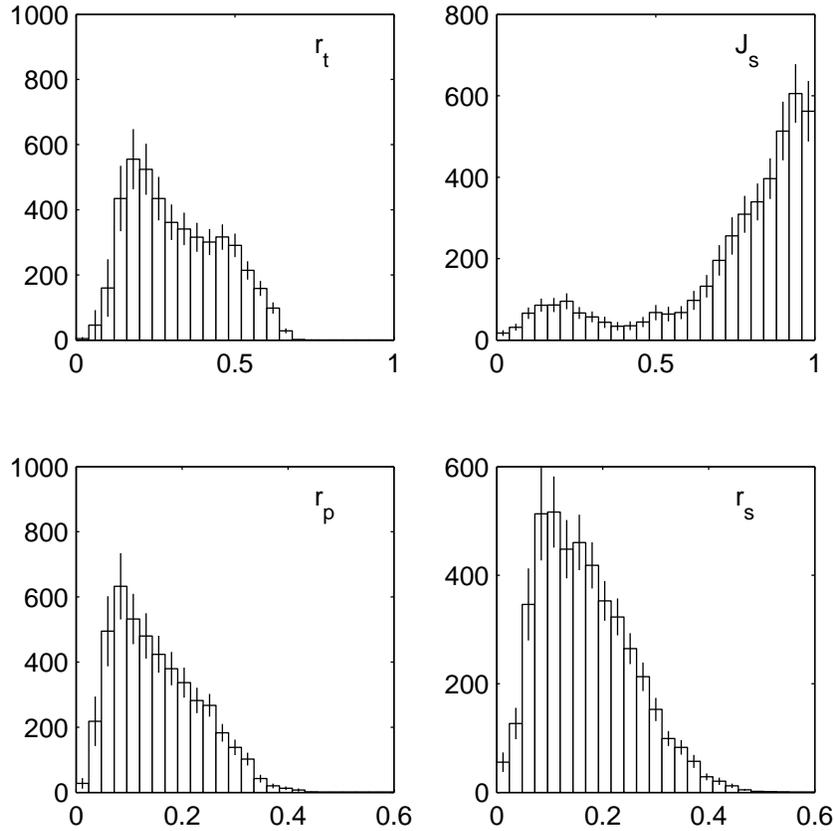}
\caption{Histogram of the elements, corrected for the observational
 selection effect}
 \label{fig:param_distr_corrected}
\end{figure*}

The histograms of the ``weighted'' binaries are shown in
Fig~\ref{fig:param_distr_corrected}. The error bars of each bin were
calculated by taking the square root of the sum of squares of the
$1/p$ values of that bin.

The histograms of the primary and secondary radii show a
dramatic rise between 0.05 and 0.1, and a moderate linear decrease
between 0.1 and 0.4. The distribution of the surface brightness ratios
has a maximum at about unity and decreases monotonically down to about
$0.5\pm0.1$; then it reverses its trend, rising to a secondary, small
maximum at about 0.2.

The histogram of the surface brightness ratios reflects the mass-ratio
distribution of the whole population of short-period binaries with
early-type primaries. This is especially true for main-sequence stars,
where for a given primary star the surface brightness ratio of a
binary is a monotonic function of its mass ratio.  Therefore, the
small local secondary maximum at 0.2 is of particular interest,
because it might be an indication for a similar local maximum in the
mass-ratio distribution.

\begin{figure*}
 \includegraphics{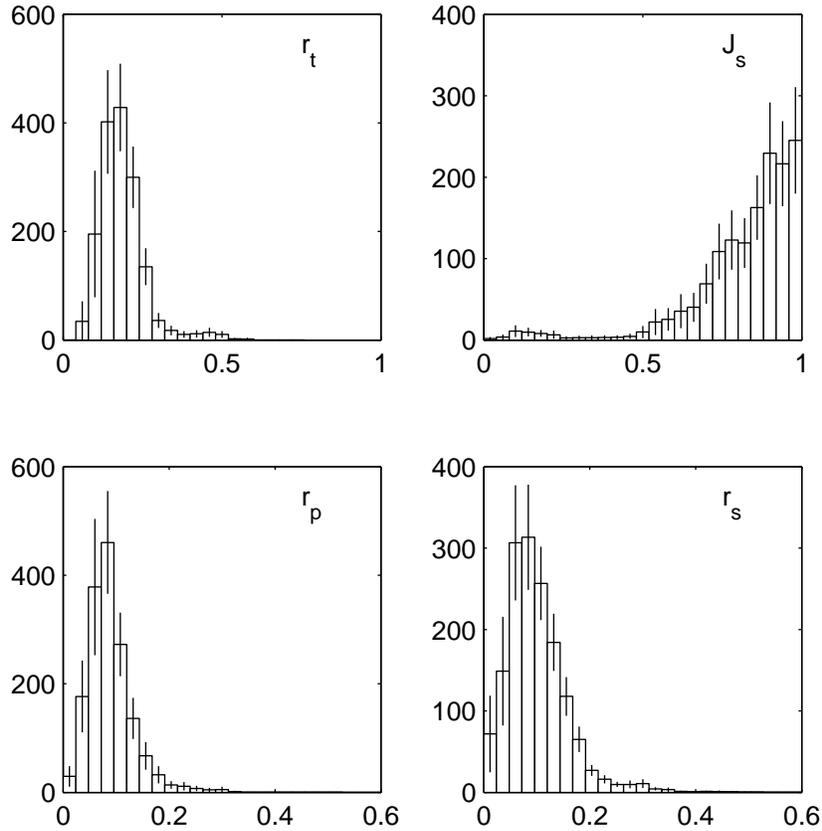}
 \caption{The same histograms as in
Fig~\ref{fig:param_distr_corrected}, for binaries with $\log P > 0.5$}
 \label{fig:param_distr_corrected_long}
\end{figure*}

However, the local maximum of the surface brightness ratio
distribution could have been caused by binaries that have gone through
mass transfer.  Such binaries should have relatively short periods. To
follow the nature of the local small maximum we plotted in
Fig~\ref{fig:param_distr_corrected_long} the corrected histograms only
for the binaries with $\log P > 0.5$.  The secondary peak at the
distribution of the surface brightness ratios almost disappears, and
the histogram decreases quite monotonically from the peak at
unity. We, therefore, suggest that the initial mass-ratio
distribution of the short-period binaries with B-type primaries rises
monotonically up to a mass ratio of unity.

\subsection{The period distribution}

\begin{figure*}
 \includegraphics{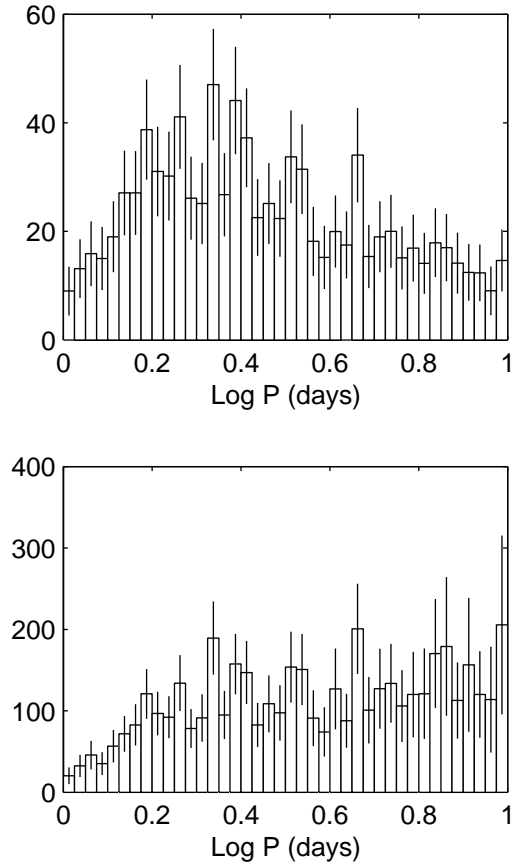}
 \caption{The period distribution of the binaries in the LMC. The top
 panel shows the period histogram of the trimmed sample, while the
 lower one shows the corrected histogram}
 \label{fig:period_distr}
\end{figure*}

To study the period distribution of the binaries in the LMC we plotted
in Fig~\ref{fig:period_distr} period histograms of the trimmed sample,
before and after the correction for the observational effects was
applied. We emphasize that if the correction was applied properly, the
lower panel represents the period distribution of all binaries in the
LMC with $I$-magnitude between 17 and 19, and with sum of radii
smaller than about 0.6, and not only the eclipsing binaries.

The corrected period histogram shows clearly a distribution that rises
up to about $\log P=0.3$, and then flattens off. However, before we
conclude that this is the actual distribution of short-period binaries
we have to consider another possible selection effect, which is
associated with the fact that the fractional length of an eclipse is
shorter for long-period binaries than for short-period binaries. This
effects turns the number of observations within an eclipse to be smaller
for long-period binaries than for short-period binaries. Therefore,
for the same two stellar components and eclipse depth, the signal
associated with the eclipse is smaller for binaries with long periods
than for those with short periods.  Consequently, the minimum
secondary radius that can be detected might be a function of the
binary period. This might reduce the number of binaries discovered
with long periods.

However, this selection effect can be substantial only for stars with
small relative radii.  Eclipsing binaries in our sample, with
secondary radii larger than 0.03 and primary ones in the range
0.1--0.2, can be detected easily for periods of 10 days. We therefore
suggest that the eclipse length effect should be quite small in
our sample.  To further check this point we calculated how an assumed
flat log distribution between 2 and 10 days would be affected by the
eclipse length effect.  To do that we took all the 143 binaries in the
sample with $0.3<\log P<0.4$ and checked their detectability if we
change their period to be 10 days and their inclination to be
$90^{\circ}$. Only eight binaries lost their detectability because of
the eclipse length effect. This means that the implication of this
effect is less than 6\%.

We therefore conclude that the period distribution of the short-period
binaries is consistent with {\it a flat log distribution between 2 and
10 days}.

\subsection{The frequency of the short-period binaries}

As stated above, the corrections we apply allow us to study the
distributions of all short-period binaries, up to 10 days, and not
only the eclipsing binaries. We wish to take advantage of this feature
of our analysis and estimate, within the limits of the sample, the
total number of binaries in the LMC and their fractional frequency.

The sum of weights of the 938 binaries in the trimmed sample is
4585. This means that we estimate there are 4585 short-period binaries
in the LMC that fulfil the constraints we have on the trimmed
sample. To get the total number of binaries with periods shorter than
10 days we have to add the high-alarm binaries and the ones with large
radii; since the added systems are close binaries, their detection
probability reaches almost 1 so we do not correct for it in their case.
When we add those we end up with 5004 binaries.

We therefore suggest that the number of binaries in the LMC, with
period shorter than 10 days, with $I$ between 17 and 19, and
for which the $V-I$ colour of the system indicates a main-sequence
primary, is about 5000. 

Our estimate is valid as far as \citet{lukas2003} detected all
eclipsing binaries and classified them as such, rather than as other
types of variables (e.g. small amplitude cepheids). We do correct for
the detection probability, including for the effect of phase of
minima, but still assumed that each binary has been correctly
identified. One could imagine that 10 to 30\% of the potentially
detectable binaries were missed for some reason, but hardly much more.
We, therefore, arbitrarily assign an error of 30\% to the number of
binaries in the sample.

We have shown that our sample consists mainly of B-type main-sequence
binaries. In order to estimate the fractional frequency of B-type
main-sequence stars which reside in binaries that could have been
detected by OGLE we have to estimate how many main-sequence {\it
single} stars were found by OGLE in the same range of magnitudes and
colours. To do that, we applied exactly the same procedure we
performed above to the whole OGLE dataset of LMC stars in the 21 OGLE
fields and found 332,297 main-sequence stars in the $I$-range of
17--19. However, the luminosity of a binary is brighter than the
luminosity of its primary by a magnitude that depends on the light
ratio of the two stars of the system. Therefore, to consider a
population of single stars which is about equal to the population of
the primaries of our binary sample we considered instead the range
between 17.5 and 19.5 $I$-magnitude, and found 705,535 stars. Taking
into account about 1\% overlap between the OGLE fields, we adopt
700,000 as a representative number of single stars in the LMC
similar to the population of our binary sample. We arbitrarily assign
an error of 50\% to this figure.

We therefore conclude that $(0.7\pm 0.4)$\% of the main-sequence
B-type stars in the LMC are found in binaries with periods shorter
than 10 days.

\begin{figure*}
 \includegraphics{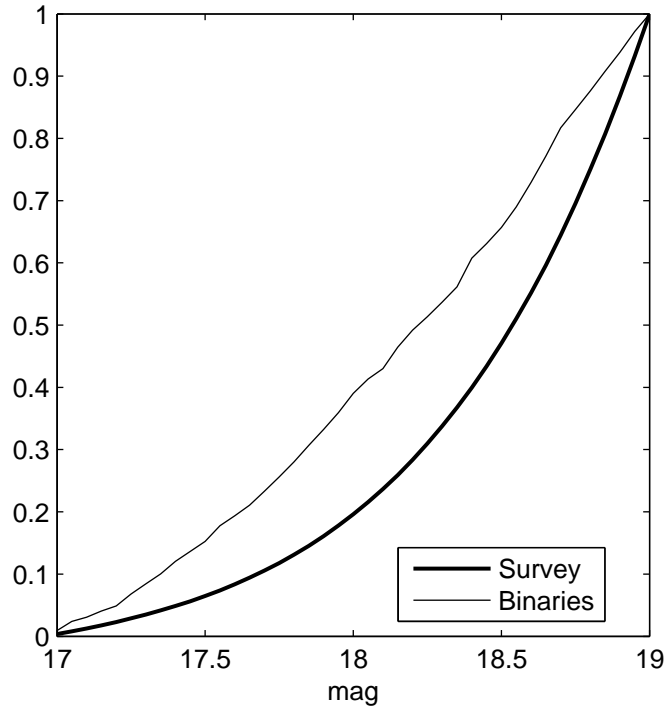}
 \caption{The magnitude cumulative distribution functions of the LMC
 binaries and the entire LMC survey. To show comparable populations
 we plot the distribution of binaries between $17$ and $19 ~I$ magnitude
 and of stars between $17.5$ and $19.5$.}
 \label{fig:mag_CDF}
\end{figure*}

A question that naturally arises is whether the binary frequency
depends on the mass of the components. While the answer to this
question is outside the scope of this study, our results allow us to
examine the variation of the binary frequency in a small magnitude
range. In Fig~\ref{fig:mag_CDF} we plot the magnitude cumulative
distribution functions of both the binaries and of the entire
survey. To display comparable populations, we plot the distribution of
binaries between $17$ and $19 ~I$ magnitude and of stars between
$17.5$ and $19.5$. While the stellar magnitude distribution rises
exponentially, the binary distribution is almost linear. Hence, the
frequency of binaries is higher for brighter stars. In fact, the
frequency of binaries in the magnitude range 17 to 18 is 1.5 percent,
while the frequency between 18 and 19 is 0.5\% --- a factor of three
smaller. It is also difficult to see how the very strong variation of
the rate of binaries with magnitude range could be credited to
detection incompleteness only, so we suggest it is real.

\subsection{The eccentricity distribution}

Another feature of the whole sample that we wish to touch upon is the
dependence of the binary eccentricity on the total radius. \citet{NZ04}
ignored the Algol-type eclipsing binaries of \citet{lukas2003}, and
considered only eclipsing binaries that did not show considerable
variation out of the eclipse. We, on the other hand, consider here all
binaries with sum of radii smaller than 0.6, and therefore, naturally,
have more systems with larger fractional averaged radii.

\begin{figure*}
 \includegraphics{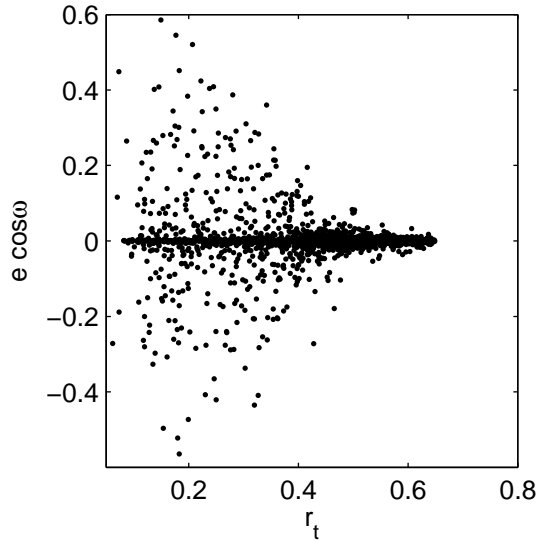}
 \caption{  $e\cos\omega$ vs. the total radius}
 \label{fig:ecosw}
\end{figure*}

Following \citet{NZ03}, Fig~\ref{fig:ecosw} shows the eccentricity,
multiplied by $\cos \omega$, as a function of the sum of radii. There
are two minor differences between our figure and that of
\citet{NZ03}. First, we present the sum of radii, while \citet{NZ03}
presented the radius, which is probably close to the average radius, as
they used a value of unity for the radius ratio, and did not allow
the ratio to change. Second, we use the value of $cos \omega$,
while \citet{NZ03} plotted the absolute value of this
parameter. Despite these minor differences, the
result obtained here is quite similar to the work of \citet{NZ04}, who found
the value of the limiting {\it averaged} radius to be 0.25.

\section{Discussion}            %
\label{discussion}

\subsection{The period distribution}

Our analysis suggests that the binaries with B-type primaries in the
LMC have flat log-period distribution between 2 and 10 days. This
distribution, which was obtain after correcting for observational
selection effects, is very different from the distributions of the
whole samples of the SMC and the LMC eclipsing binaries
\citep{udalski1998, lukas2003}.  Evidently, correcting for the
selection effects reveals a substantially different period
distribution.

The period distribution derived here is probably not consistent with
the period distribution of \citet{DM91}, who adopted a Gaussian with
$\overline{\log P}=4.8$ and $\sigma=2.3$, $P$ being measured in days.
Their distribution would provide within a log P interval almost twice
as many systems at $\log P = 1.0$ than at $\log P = 0.3$ (more
precisely, the factor would be 1.7), while the new derived
distribution is probably flat.

It is interesting to note that \citet{heacox1998} reanalysed the data
of \citet{DM91} and claimed that $f(a)$, the distribution of the
G-dwarf orbital semi-major axis, $a$, is $f(a) \propto a^{-1}da$,
which implies a flat log orbital separation distribution. This is
equivalent to the present probable result, although the latter refers
to LMC binaries with B-type primaries, and is limited only to a very
small range of orbital separation. On the other hand,
\citet[][hereafter HaMUA03]{halbwachs2003} analysed the spectroscopic
binaries found within the G- and K-stars in the solar neighbourhood
and the ones found in the Pleiades and Praesepe, and found that the
log-period distribution is ``clearly rising until about 10 days''.

However, HaMUA03's sample included only a few systems with periods
longer than 2 and shorter than 10 days. Actually, there were 7 such
systems in the ``extended sample'' of nearby stars and 5 additional
binaries in the Pleiades and Praesepe. In fact, if we divide these 12
binaries into two bins with equal log width we find 6 in the
short-period bin and 6 in the long-period one. It is true that the
correction factor of the spectroscopic binaries of these two bins is
somewhat different, and therefore the distribution of this sample
might still be consistent with the conclusion of HaMUA03. However, the
small number of binaries with periods shorter than 10 days in the G-
and K-binaries is also consistent with flat log distribution below 10
days. We verified that by building the cumulative distribution of the
$\log P$ values of the 12 G- and K-binaries, and comparing it with the
theoretical flat log distribution. The one-sided
Kolmogorov-Smirnov test gives a probability of 0.7 that the
observed distribution is consistent with the flat log distribution.

The flat log-period distribution implies a flat log
orbital-separation distribution for a given total binary mass. Such a
distribution indicates that there is no preferred length scale for the
formation of short-period binaries \citep{heacox1998}, at least in the
range between 0.05 and 0.16 AU, for a total binary mass of 5
$M_{\odot}$. Alternatively, the results indicate that the specific
angular momentum distribution is flat on a log scale at
the $10^{19}$ $cm^2s^{-1}$ range.

A note of caution is in order here, as the present analysis refers to
{\it all} eclipsing binaries in a given absolute magnitude range.
Even if the overwhelming majority of the binaries lie on or close to
the main sequence, a small unknown number of them may have gone
through a mass exchange phase, which would have affected both the
orbital period and the stellar radii. This is very important for the
interpretation of the new result, since the most interesting quantity
is probably the {\it primordial} period distribution, which was not
affected by the subsequent binary evolution.  In fact,
\citet{harries2003} and \citet[][hereafter HiHoHa05]{hilditch2005}
studied in details 50 eclipsing binaries in the SMC and found more
than half of them to be in a semi-detached post-mass-transfer state.
However, the binaries studied in the SMC, with limiting brightness of
$B < 16$, are brighter than the systems in the trimmed sample, and are
composed mainly of O to B1-type primaries, while most of the primaries
in the present sample are cooler B-type stars. Apparently, the
fractional radii of the stars studied by HiHoHa05 is substantially
larger than the typical fractional radii of the present sample. We
therefore suggest that while a large fraction of the HiHoHa05
sample has gone through mass exchange, most of the binaries studied here
are still on the main sequence.

We therefore suggest that the short-period B-type binaries in the LMC
might have had a primordial flat log-period distribution between 2 and
10 days, and a similar feature {\it could} have been found in binaries
with G-type primaries in the solar neighbourhood. Although this
probable result refers to a very narrow period range, a binary
formation model should account for this scale-free distribution, which
might be common in short-period binaries.

\subsection{The binary fraction}

It would be interesting to compare the frequency of B-star binaries
found here for the LMC with a similar frequency study of B stars in
our Galaxy. However, such a large systematic study of eclipsing
binaries is not available. Instead, a few radial-velocity and
photometric searches for binaries in relatively small Galactic samples
were performed. In what follows we compare the frequency derived here
with results of radial-velocity searches for binaries, in samples of
Galactic B stars as well as within the nearby K and G stars. In
addition, we discuss the binary frequency found by the HST photometric
monitoring of 47 Tuc.

\subsubsection{Radial-velocity searches for binaries}

\citet{wolff78} studied the frequency of binaries in sharp-lined
B7--B9 stars.  Her Table~1 presents 73 such stars with $V\sin i <
100$~km\,s$^{-1}$, among which 17 are members of binary systems with
$P_{\mathrm orb} < 10$~days. This represents a percentage of $23$\%,
much larger than the frequency we find in the LMC.  On the one hand,
the frequency of \citet{wolff78} may be considered a lower limit,
since her sample was biased toward sharp-lined stars, because her
purpose was to determine what fraction of the intrinsically slowly
rotating B stars are members of close binary systems. This introduced
a bias toward low $V\sin i$ values and therefore small orbital
inclinations, hence lower detection probabilities, assuming the spin
and orbital axes are aligned. On the other hand, the sample is
magnitude limited, but the correction for the Branch bias
\citep{branch76} was not applied. The most extreme correction factor
for the bias towards the more luminous binary systems, which holds for
systems with two identical components, is $f=0.35$. Applying this
factor to the rate of late B-type binaries with $P_{\mathrm orb} <
10$~days leaves a minimum frequency of $8\pm 2$\%.  This remains one
order of magnitude larger than our frequency in the LMC.

Similar and even higher values of binary frequency were derived for B
stars in Galactic clusters and associations.  \citet{ML91} have
determined the rate of binaries among B stars (a few O and A-type
stars are also included in the sample) in the Orion OB1 association,
and obtained an overall frequency of binaries with $P_{\mathrm orb} <
10$~days of $26\pm 6$\%. There is a slight trend toward a higher rate
of binaries among the early B stars than among the late ones. For B4
and earlier stars they found 60\% binaries, while only $\sim 27$\% of
the B5 and later stars were found to be RV variables, which is
reminiscent of what we find in the LMC. However, this is hardly
significant as the latter frequency is based only on 3 variables.

\citet{GM01} found a rate of binaries as high as $82$\% among
stars hotter than B1.5 in the very young open cluster NGC~6231, though
in a sample of 34 stars only. Most orbits have periods shorter than 10
days.  \citet{R96} estimated a rate of binaries of at least $52$\% for
36 B1-B9 stars in the same cluster, but his data did not allow him to
determine the orbital periods.

These studies show that binaries hosting B-type stars tend to be
more frequent in young Galactic clusters than in the field. However,
even though our LMC binaries are representative of the LMC field
rather than of LMC clusters, the frequency we derived appears much
lower than that of the binaries in the Galactic field.

One survey of a complete, volume limited, sample of binaries in the
solar neighbourhood is that of HaMUA03, who carefully studied a
complete sample of K- and G-type stars, and not B stars.  Out of a
bias free sample of 405 in the solar neighbourhood, they found 52
binaries, 11 of them with periods shorter than 10 days. This results
in a fractional frequency of $(2.7\pm0.8)$\% binaries, if one takes
just the Poisson distribution to estimate the error.  We therefore
suggest that the fractional frequency of K- and G-type stars in the
solar neighbourhood found in binaries with periods shorter than 10
days is higher by a factor of about $3.9\pm1.9$ than the corresponding
frequency of B-type stars in the LMC. The error estimate of the factor
between the two fractional frequencies was derived by assuming normal
distribution, and is only a coarse estimate, because of the small
number statistics of the K- and G-type binaries. In fact, the
difference between the two fractional frequencies is at about the
$2\sigma$ level.

\subsubsection{HST Photometric monitoring of 47 Tuc}

A few photometric studies were devoted to the search of eclipsing
binaries in Galactic globular clusters \citep[e.g.,][]{yan96,
kaluzny97}. One of the recent photometric studies is the HST 8.3-day
observations of 47 Tuc \citep{albrow01}, which monitored 46,422 stars in
the central part of the cluster and discovered 5 eclipsing binaries
with periods longer than about 4 days. Assuming a flat log-period
distribution of binaries, \citet{albrow01} concluded that the
primordial binary frequecy up to 50 years was $(13\pm6)$\%. This
estimate translates into $(2\pm1)$\% for periods between 2 and 10
days. This value, derived for late-type stars, is much smaller than
the frequency derived by the radial-velocity surveys of B stars, and
is close to the $(0.7\pm0.4)$\% frequency we derived for the LMC.

\subsubsection{Comparison between the two approaches}
 
We conclude that the frequency of binaries we find in the LMC is
substantially smaller than the frequency found by Galactic
radial-velocity surveys. The detected frequency of B-type binaries is
larger by a factor of 10 or more than the frequency we find, while the
binary frequency of K- and G-type stars is probably larger by a factor
of four. On the other hand, the binary frequency found by photometric
searches in 47 Tuc is only slightly higher and still consistent with
the frequency we deduced for the LMC. It seems that the frequency
derived from photometric searches is consistently smaller than the one
found by radial-velocity observations.

We are not aware of any observational effect that could cause such a
large difference between the radial velocity and the photometric
studies.  The sensitivity of the two approaches depends on the ability
to discover binaries with secondaries of small radii, in the
photometry case, and small masses, in the case of the radial-velocity
searches.  In order to account for the large differences between the
photometric and the spectroscopic searches we have to assume that the
photometric searches detect only ten percent of the binaries and
missed all the others, that could have been discovered by
radial-velocity techniques. Although \citet{wolff78} suggested that
many short-period binaries in her radial-velocity sample have low-mass
secondaries, this effect can not explain a factor of ten in the
detected frequency. The discussion of the radii distribution of our
sample also indicates that if such an effect is present in the LMC
binaries it must be small. After all, the radial-velocity searches are
not sensitive to small-mass secondaries, and they too suffer from
selection effects which depends on their precision. Therefore, the
large difference in the binary frequencies is probably real and
remains a mystery. Obviously, it would be extremely useful and
interesting to have studies of eclipsing binaries similar to the
present one in our Galaxy, as well as in other nearby galaxies.

\subsection{The stellar radii and the surface brightness ratios}

The distributions of the two fractional radii depend strongly on the
derivation of the ratio of radii, which is not well determined from
the lightcurves. Still, a statistical analysis of the distributions is
meaningful, given the high correlation between the true and derived
$k$ values, as shown in the simulations in Paper I.  The distributions
of the radii show sharp peaks at the same value of 0.1, as can be seen
in Fig~\ref{fig:param_distr_corrected}, and in
Fig~\ref{fig:param_distr_corrected_long} in particular. This suggests,
but admittedly not prove, that our sample is mostly composed of stars
with similar radii. The distribution of $J_s$ tends to reinforce this
interpretation, since it suggests a mass ratio --- hence a ratio of
radii --- close to one.  

For the primary, the rise between $0.05$ and $0.1$ is due essentially
to the limit imposed on the orbital period and to the minimum radius
of a star on the ZAMS. On the other hand, the sharp rise of the
secondary radius distribution up to a radius of about 0.1 is probably
real. This feature of the secondary radius distribution could have
been the result of a selection effect --- too small a secondary results
in an eclipse too shallow to be detected by the OGLE photometry. We
suggest that this is not the case. The detectability limit of the
sample is such that any binary with eclipse deeper than about 0.1 mag,
corresponding to 10\% drop in the system brightness, is included in
the sample. Such a drop can be caused by a secondary with a radius
which is about 0.3 of the primary radius, for inclination of
$90^{\circ}$ and $J_s < 1$. If the primary radius peaks at $\sim 0.1$,
the sample is then complete for secondary radius down to about
0.03.

The peaks at about 0.1 in Fig~\ref{fig:param_distr_corrected_long}
indicate that a substantial fraction of the primary and the secondary
stars in the sample have radius of 2.5--3 $R_{\odot}$, assuming a
typical period of 7 days in Fig~\ref{fig:param_distr_corrected_long}
and a typical mass of 5 $M_{\odot}$. The sharpness of the peaks is
probably caused by the magnitude limits of the trimmed sample.

\subsection{The eccentricity distribution}

Fig~\ref{fig:ecosw}, with the eccentricity as a function of the sum of
the two radii, displays upper and lower envelopes that can be
approximated with two straight lines, with a slope of about 1.7 in
absolute value. Although both the shape and slope of these envelopes
may be plagued by selection biases, it is interesting to remark that
the populated area corresponds to the condition
\begin{equation}
$$r_t=r_p+r_s\lesssim
0.59\,(0.85-|e\cos\omega|) \ \ .
\end{equation}
This roughly generalizes the
condition $r_{average}\lesssim 0.25$ to the case of eccentric systems,
which cannot keep an eccentricity larger than some limiting
value. This limiting value of the eccentricity depends on the stellar
radii, whose sum can not be larger than a given critical fraction of
the distance {\it at periastron}. Because of the $\cos\omega$ factor,
it is reasonable to assume that the real relation is rather
$$r_p+r_s\lesssim 0.5\,(1-e) \ .$$
Systems with eccentricities larger than that will undergo partial
tidal circularization until they satisfy this inequality; after that,
tidal effects become almost inefficient. The boundary is clear-cut
even though the systems do not share the same age, due to the very steep
dependence of circularization time on relative radius: $t_{\rm circ}\propto
(R/a)^{-21/2}$ \citep[][]{zahn75, zahn77}.

Note also that there are quite a few binaries with small radii for
which the parameter $e \cos \omega$ is very close to zero. It is
difficult to assume that all these system have $\cos
\omega=0$. Probably those systems have circular orbits. It would be
interesting to find out whether these systems were circularized during
their main-sequence phase, despite their small radii, or maybe their
very small eccentricity is primordial.

\section{Summary}               %
\label{summary}

The OGLE data set has allowed us to analyze for the first time the
statistical characteristics of short-period binaries with B-type
primaries of an entire galaxy.  We analyzed the 2580 eclipsing LMC
lightcurves found by \citet{lukas2003} and obtained orbital elements
for 1931 systems. We further derived some statistical features of a
trimmed sample of 938 binaries with periods shorter than 10 days,
main-sequence primaries, and $I$-magnitude between 17 and 19.  After
correcting for the inclination effect, we find that the log-period
distribution of the short-period binaries is probably flat between 2
and 10 days. Finally, we suggest that $(0.7\pm 0.4)$\% of the
main-sequence B-type stars in the LMC are found in binaries with
periods shorter than 10 days. This is substantially smaller
than the fraction found in the nearby K- and G-type stars and much
smaller than that of nearby B stars. We are
preparing another paper on the distribution of stars in the SMC, to
find out the metallicity effect of the frequency of binaries.

This work is a further step in the study of the extragalactic
binaries. Most previous studies concentrated on individual systems
(see a review by \citet{ribas2004} and references therein). The
photometry of a whole galaxy, the LMC in our case, can provide
photometry of thousands of eclipsing binaries, together with distance
modulus which is common to all binaries in the same galaxy. The known
distances to the galaxies in the local group can provide us with
absolute magnitude for a large set of early-type eclipsing binaries
that is not available for Galactic binaries. Therefore, the
statistical analysis of the extragalactic binaries presents great
potential to our understanding of the formation of early-type binaries
and stellar models.

\section*{Acknowledgments}
We are grateful to the OGLE team, and to L. Wyrzykowski in particular,
for the excellent photometric data set and the eclipsing binary
analysis that was available to us. We thank J. Devor, G. Torres and I.
Ribas for very useful comments. The remarks and suggestions of the
referee, T. Zwitter, helped us to substantially improve the algorithm
and the paper. This work was supported by the Israeli Science
Foundation through grant no. 03/233.

\bibliographystyle{mn2e}
\bibliography{ref}
\end{document}